# Mechanically and electrically switchable triferroic altermagnet in a pentagonal $FeO_2$ monolayer

Deping Guo[1,#,*], Jiaqi Dai[2,3,#], Renhong Wang[2,3], Cong Wang[2,3*] and Wei Ji[2,3,*]

[1]*College of Physics and Electronic Engineering, Center for Computational Sciences, Sichuan Normal University, Chengdu, 610101, China*

[2]*Beijing Key Laboratory of Optoelectronic Functional Materials & Micro-Nano Devices, School of Physics, Renmin University of China, Beijing 100872, China*

[3] *Key Laboratory of Quantum State Construction and Manipulation (Ministry of Education), Renmin University of China, Beijing, 100872, China*

[*]*Corresponding authors.* Email: dpguo@sicnu.edu.cn (D.G.), wcphys@ruc.edu.cn (C.W.), wji@ruc.edu.cn (W.J.).

## Abstract

Two-dimensional multiferroics promise low-power, multifunctional devices, yet the intrinsic coexistence and mutual control of three coupled ferroic orders in a single layer remains elusive. Here, we identify pentagonal monolayer $FeO_2$ as an intrinsic triferroic altermagnet where ferroelectric (FE), ferroelastic (FA), and altermagnetic (AM) orders coexist and tightly coupled, accompanied by a competing antiferroelectric (AFE) phase using first-principles calculations. The solely presence of glide mirror $M_x$ symmetry in a $FeO_2$ sublayer, with the breaking of four-fold rotation $C_{4z}$ symmetry, induces in-plane vector ferroelectricity and twin-related ferroelastic strains. Both FE and AFE phases break combined parity–time symmetry and display sizable altermagnetic spin splitting with Néel temperatures over 200 K. Electric-field induced rotation of the FE polarization reverses the sign of the spin splitting, while in-plane uniaxial strain triggers ferroelastic switching that simultaneously rotates the FE polarization vector by 90° and reverses the AM state. These electric-field- and strain-mediated pathways interlink six distinct polarization states that can be selected purely by electric fields and/or mechanical strain. This work extends intrinsic triferroicity to pentagonal monolayers and outlines a symmetry-based route toward mechanically and electrically configurable altermagnetic spintronics.

The pursuit of two-dimensional (2D) intrinsic multiferroic (MF) materials, characterized by the coexistence and cross-coupling of multiple ferroic orders, has been strongly motivated by their potential to realize mutual control among multiple polarization states [1,2]. This capability is crucial for developing minimized, low-power-consuming, non-volatile, and high-speed multifunctional devices [3]. While identifying materials that simultaneously host coupled ferroic orders, such as ferroelectricity (FE) and ferromagnetism (FM), remains challenging [4], anti-ferromagnetic (AFM) orders are frequently found to coexist and couple with other anti-ferroic or ferroic states [5,6]. However, conventional AFM materials, lacking net magnetization and spin-split bands, significantly limit their application in information storage and processing devices.

Altermagnetism (AM), a unconventional antiferromagnetism discovered recently, addresses this limitation by exhibiting distinct spin splitting in momentum space [7,8]. Altermagnetism arises from nonrelativistic spin-group symmetries [9], in which sublattices with opposite spins are related by crystallographic rotation or mirror operations, rather than by inversion or translation symmetry. This results in the breaking of combined parity–time (PT) symmetry or time-reversal–translation (Tτ) symmetry. This intrinsic spin splitting effectively provides spin polarization within anti-ferromagnets, significantly broadening the search for multiferroic materials into the realm of AFM materials suitable for device applications. Recently studies have been demonstrated that the sign of altermagnetic spin splitting ($S$=$E_{up}$ - $E_{down}$) can be reversed by switching ferroelectric polarization in ferroelectric altermagnets [10–18]. Ferroelasticity (FA), the mechanical analogue of FE and FM, enables significant reversible shape changes with hysteresis. Remarkably, ferroelastic domain switching can also invert the sign of $S$ through the altermagneto-elastic effect [19]. These recent advances raise an intriguing question: can FE, FA, and AM intrinsically coexist within the same 2D material? Such coexistence, if realized, would enable non-volatile and multi-states switching of altermagnetic properties using electric fields and/or mechanical strain.

Pentagonal 2D materials were first theoretically predicted with the introduction of penta-carbon, a carbon allotrope [20]. Since then, numerous pentagonal 2D structures, exhibiting significantly lower symmetry compared to traditional 2D layers like graphene or 1T- or 2H-phase transition metal dichalcogenides, have been identified in both non-magnetic and magnetic systems [21–23]. These pentagonal 2D layer exhibit FE [24], piezoelectricity [25,26], or FA [27], arising primarily due to the absence of inversion symmetry or pronounced lattice anisotropy. Experimental synthesis and verification of stability have been achieved for several pentagonal 2D materials, such as $PdSe_2$ [28,29], $PdTe_2$ [30] and PdPSe [31], confirming their feasibility and thermal robustness. These distinctive features make pentagonal 2D materials promising candidates for simultaneously hosting ferroelastic and ferroelectric orders. A crucial remaining question is thus whether any of these pentagonal 2D candidates can also exhibit AM.

In this work, we theoretically predict the intrinsic coexistence of AM, (anti)FE, and FA within a pentagonal $FeO_2$ (p-$FeO_2$) monolayer. Using density functional theory (DFT) calculations, we found a FE and an antiferroelectric (AFE) phase in the p-$FeO_2$ monolayer and further verified their dynamic stability through phonon spectra calculations. Subsequent symmetry analysis reveals that they both break parity-time symmetry and exhibit pronounced altermagnetic spin splitting, with theoretical magnetic transition temperatures exceeding 200 K. Furthermore, we demonstrate that the altermagnetic spin splitting can be tuned via in-plane polarization switching, sublayer displacement, and strain-induced lattice anisotropy. This lattice anisotropy, combined with applied in-plane strain, establishes FA, enabling reorientation of polarization directions and consequently reversing the sign of altermagnetic spin splitting. These intertwined polarization states result in six distinct, electrically and/or mechanically switchable polarization configurations of the p-$FeO_2$ monolayer. Our findings expand the family of 2D intrinsic triferroic materials and highlight pentagonal 2D layers as promising candidates for further multifunctional device design.

Our density functional theory (DFT) calculations were carried out using the generalized gradient approximation for the exchange-correlation potential [32], the projector augmented wave method [33] and a plane-wave basis set as implemented in the Vienna ab-initio simulation package (VASP) [34,35]. In all calculations, the Grimme's D3 form vdW correction was applied to the Perdew Burke Ernzerhof (PBE) exchange functional (PBE-D3) [36]. Kinetic energy cut-offs of 700 eV and 500 eV for the plane wave basis set were used in structural relaxations and electronic calculations, respectively. All atomic positions and lattices were fully relaxed until the residual force per atom was less than 0.01 eV/Å. An 8×8×1 *k*-mesh was adopted to sample the Brillouin zone. A vacuum layer, over 15 Å in thickness, was used to reduce interactions among image slabs. On-site Coulomb interactions on the Fe (effective $U$= 3.0 eV) $d$ orbitals were considered using a DFT+U method [37]. Phonon spectrums were calculated using the density functional perturbation theory, as implemented in the PHONOPY code [38]. In phonon spectra calculations, the dispersion correction was made at the van der Waals density functional (vdW-DF) level [39] with the optB86b functional for the exchange potential (optB86b-vdW) [40]. The electric polarizations were derived using the Berry phase method [41]. The value of polarization is obtained by calculating the difference between the $+P$ state and the $-P$ state, as expressed by $P = \frac{+P-(-P)}{2}$. The multiferroic transition barrier was estimated by using the nudged elastic band method [42].

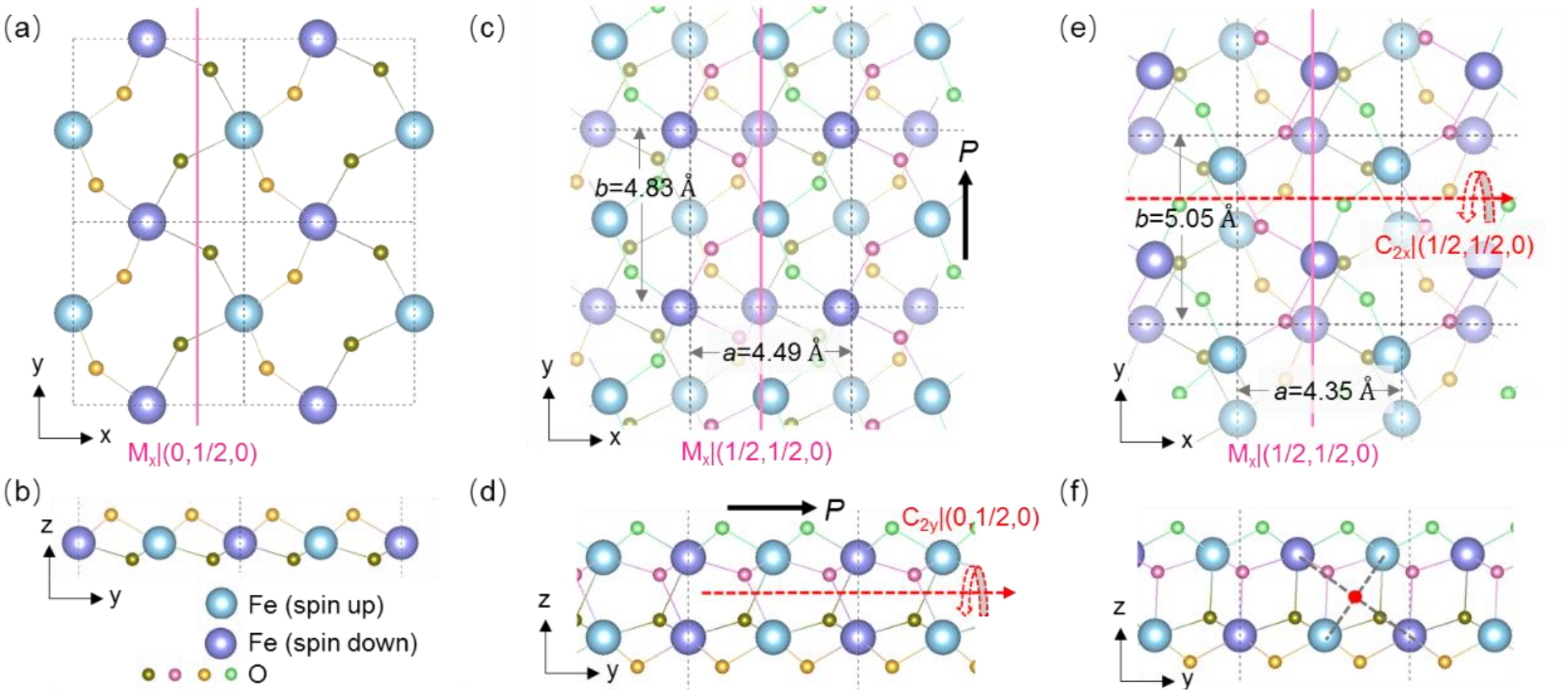

FIG. 1. (a) Top and (b) side views of one sublayer of p-$FeO_2$ monolayer. (c) Top and (d) side views

of p-$FeO_2$ monolayer in the FE phase. The black arrow denotes the direction of polarization. (e) Top and (f) side views of p-$FeO_2$ monolayer in the AFE phase. The pink lines and red arrows illustrate mirror symmetry and rotation symmetry, respectively. The red dot denotes the inversion center.

A p-$FeO_2$ monolayer contains two sublayers. Each sublayer consists of a layer of magnetic atom sandwiched between two oxygen layers, as shown in Figs. 1(a) and 1(b). Two sublayers covalently bond to each other [Figs. 1(c)-1(f)], adopting a structure analogous to that of $PdSe_2$ [28,29] and $VTe_2$ [43]. Each sublayer only preserves a mirror symmetry operation $\{M_x | (0, 1/2, 0)\}$ [pink line in Fig. 1(a)], implying the possibility of spontaneous electric polarization along either the *y*- or *z*-axis. Stacking the two sublayers into a monolayer with space group $Pca2_1$ (No. 29) gives rise to a glide mirror symmetry operation $\{M_z | (1/2, 0, 0)\}$. This operation eliminates likely electric polarization along the *z*-axis, but preserves the electric polarization directions of the two sublayers being parallel along the *y*-axis, thereby resulting in a FE phase [Figs. 1(c) and 1(d)]. The FE-phase monolayer exhibits a spontaneous in-plane polarization of $P_y$ = 12.5 μC/cm$^2$ along the *y*-axis. This value is comparable to those theoretically predicted for prominent 2D ferroelectrics, such as SnTe (22 μC/cm$^2$) [44], GeTe (32.8 μC/cm$^2$) [45], and $CuCrSe_2$ (19 μC/cm$^2$) [46]. Lateral shifts of the two Fe atoms in the top sublayer of the FE phase, together with the corresponding relaxations of adjacent O atom, form of an AFE structure. This structure belongs to space group $P2_1/c$ (No. 14), exhibiting inversion symmetry [Figs. 1(e) and 1(f)]. Consequently, the polarizations of the two sublayers are oppositely aligned and cancel each other, with the bottom sublayer carrying a polarization of (0, 8.8, -0.9) μC/cm$^2$.

By evaluating four different magnetic configurations for the FE p-$FeO_2$ monolayer (Fig. S1 [47]), we found that the AFM1 configuration [Figs. 1(c) and 1(d)] is at least 10 meV/ Fe more stable than any of other configurations. In this magnetic configuration, the optimized lattice constants are $a$=4.49 Å and $b$=4.83 Å. Nevertheless, the atomic displacements associated with the transition from the FE to AFE phase alter the lowest energy magnetic configuration from AFM1 [Figs. 1(c) and 1(d)] to AFM2 [Figs. 1(e) and 1(f), Fig. S2 [47]], exhibiting a shrank $a$=4.35 Å and an expanded $b$=5.04 Å. Magnetic anisotropy mapping reveals that the easy axis of magnetization in both phases is oriented along the $z$-axis (Fig. S3 [47]), with single-ion anisotropy energies of 0.29 meV/Fe for the FE phase and 0.35 meV/Fe for the AFE phase. An anisotropic Heisenberg model that includes the first, second, third and fourth-nearest couplings and the single-ion anisotropy was employed in Metropolis Monte Carlo simulations to estimate the magnetic transition temperatures (Fig. S4 and Supplementary Notes [47]). The estimated Néel temperatures are 230 and 202 K for the FE and AFE phases, respectively. Phonon dispersion calculations and ab initio molecular dynamics simulations confirm the dynamic and thermal stability of the both phases (Fig. S5 [47]). The two sublattices with opposite spins in both the FE and AFE phase are related by a mirror operation and a two-fold rotation. They share the same mirror symmetry $\{M_x \mid (1/2, 1/2, 0)\}$, as indicated by the pink lines marked in Figs. 1(c) and 1(e), but differ in their rotational operations, namely $\{C_{2y} \mid (0, 1/2, 0)\}$ for FE [red arrow in Fig. 1(d)] and $\{C_{2x} \mid (1/2, 1/2, 0)\}$ for AFE [red arrows in Fig 1(e)]. The fact that the opposite spins are connected exclusively by $C_2$ rotations and the $M_x$ mirror operation indicate that the joint parity and time (P-T) symmetry is broken in both phases, resulting in altermagnetic spin splitting states.

The band structure of the FE p-$FeO_2$ monolayer [Fig. 2(a)] exhibits pronounced spin splitting of several hundred meV in path M-Γ-M′, along with a band gap of 0.51 eV, identifying the monolayer as an altermagnetic semiconductor. Explicit spin splitting also presents in the bandstructures of the AFE p-$FeO_2$ monolayer, especially in the valence bands, accompanying with a direct bandgap of 0.42 eV [Fig. 2(b)]. We introduce $S_{\mathrm{FE}} = E_{\mathrm{FE}}^{\downarrow} - E_{\mathrm{FE}}^{\uparrow}$ to denote the spin splitting characteristics along the Γ-M path for the second conduction band [CBM +1, black arrow in Fig. 2(a)] in the FE phase. Similarly, we define $S_{\mathrm{AFE}} = E_{\mathrm{AFE}}^{\uparrow} - E_{\mathrm{AFE}}^{\downarrow}$ to describe the spin splitting along the Γ–M path for the valence band maximum (VBM) in the AFE phase, as indicated by the black arrow in Fig. 2(b). In either phase, the distribution of $S_{\mathrm{FE}}$ or $S_{\mathrm{AFE}}$ is four-fold and exhibit a $C_{2z}$ rotation symmetry (Fig. S6 [47]). The spin-splitting mappings reveal maximum values of ~396 meV in the FE phase and ~230 meV in the AFE phase. Although smaller than those of certain 2D altermagnets with piezomagnetism (~1 eV) [48,49], these values exceed or comparable to those reported for $RuF_4$ (~240 meV) [50] and other 2D altermagnets [51–55]. When $S_{FE}$ or $S_{AFE}$ is positive (negative), the first Brillouin zone (BZ) region containing path Γ-M (gray dashed line) is colored orange or pink (green or blue) in the inset of Fig. 2(a) or 2(b).

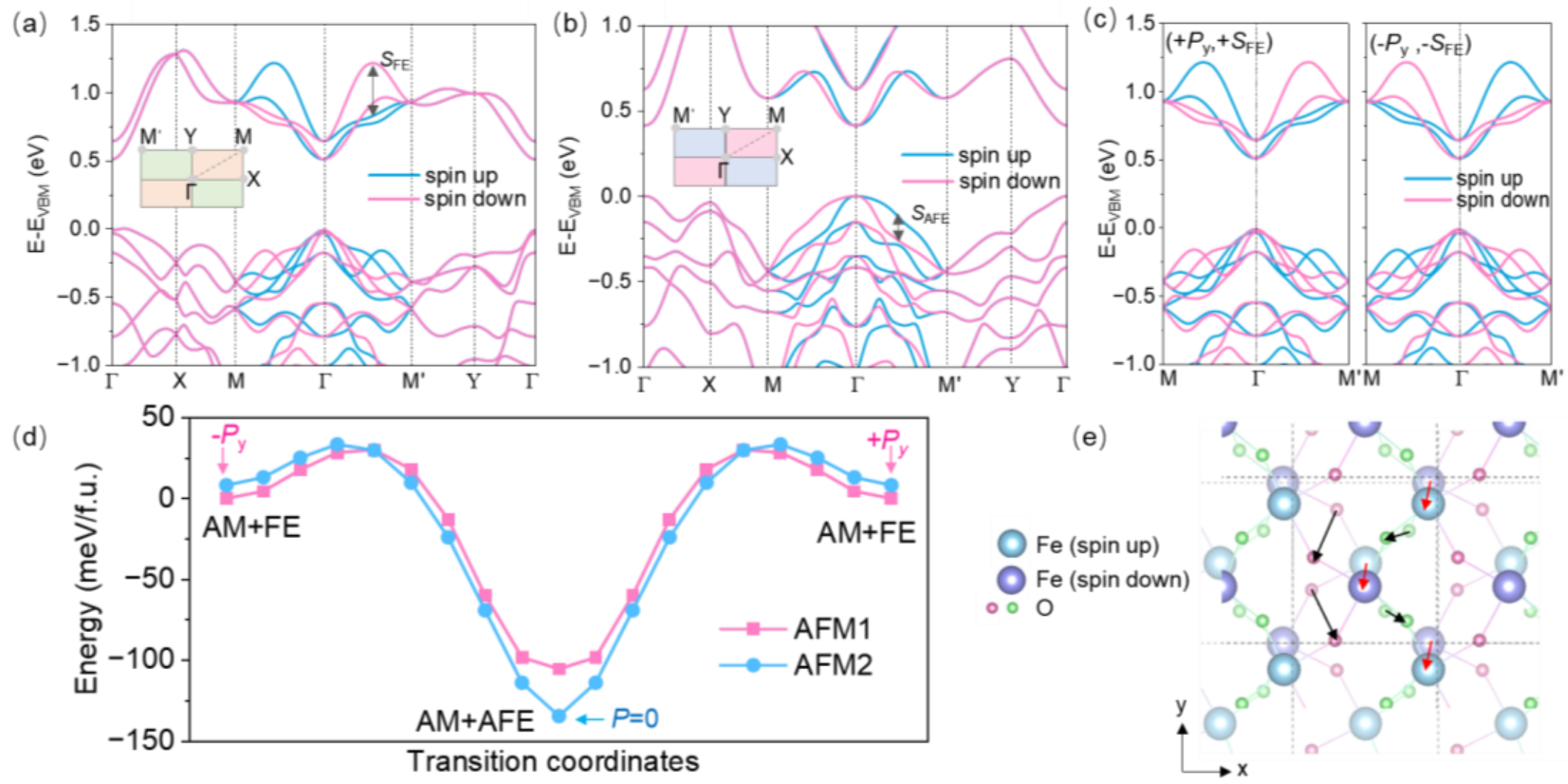

FIG. 2. (a)-(b) Band structure of the p-$FeO_2$ in the (a) FE phase and (b) AFE phase. The illustration shows the high-symmetry path and spin splitting symmetry in the Brillouin zone. The orange regions represent the distribution of spin splitting values in the FE phase, denoted as $+S_{FE}$, while the green regions indicate the opposite distribution, denoted as $–S_{FE}$. In the AFE phase, the distributions $+S_{AFE}$ and $–S_{AFE}$ are represented by pink and blue regions, respectively. (c) Band structures of FE p-$FeO_2$ under different polarization directions. (d) Ferroelectric transition barriers between AM+FE and AM+AFE states. The pink curve illustrates the transition pathway where the AFM1 configuration was set initially, while the blue curve indicates the pathway where AFM2 was initially set in all images along the path. (e) The schematic diagram illustrates the relative displacements between the top sublayers in the FE and AFE phases, where the red arrows indicate the displacement of Fe atoms and the black arrows represent the displacement of O atoms. Lighter-colored balls represent atoms in the top sublayer in the FE phase, whereas darker-colored balls correspond to atoms in the top sublayer in the AFE phase.

As stacking of two sublayers modulates the in-plane ferroelectric polarization, the monolayer exhibits three distinct polarization ($P$) states: $+P_y$, 0, and $-P_y$ [Fig. 2(c)]. These three electric polarization states are associated with two types of altermagnetic bandstructures shown in Figs. 2(a) and 2(b). The switching of $P_y$ accompanies a magnetic transition from AFM1 to AFM2. However, NEB calculations cannot determine the occurring geometry for the magnetic transition. Thus, both magnetic configurations were considered in two separate NEB calculations for estimating the switching barrier and pathway, which yield comparable results. As indicated by Fig. 2(d), the AFE phase is energetically more favorable, with a transition barrier of 168 meV/f.u. (33 meV/f.u.) from the AFE (FE) to the FE (AFE) phase in the AFM-2 (AFM-1) configuration. During the ferroelectric phase transition (Fig. S7 [47]), the bottom sublayer largely retains its relative positions, while the top-sublayer atoms undergo pronounced in-plane displacements. The relative displacements of the top sublayers between the FE and AFE phases are presented in Fig. 2(e).These barriers are slightly smaller than, but comparable to, those of monolayer $CuWP_2S_6$ [0.33 eV/f.u. (0.15 eV/f.u.)], in which the AFE phase also exhibits better stability than the FE phase [11]. Notably, the spin splitting also reverses in response to the flipped electric polarization direction [Fig. 2(c)], which can be triggered by external electric fields. Accordingly, the three polarization states can be represented as ($+P_y$, $+S_{FE}$) [Fig. 2(d) left], (0,$+S_{AFE}$), and ($-P_y$,$-S_{FE}$) [Fig. 2(c) right].

Although the AFE phase is energetically favorable in the freestanding case, in-plane epitaxial strain can effectively engineer the relative stability of the FE and AFE phases (Fig. S8 [47]). The in-plane strain can even stabilize the p-$FeO_2$ monolayer relative to the experimentally synthesized 1T-$FeO_2$ [56], as detailly discussed in Fig. S9 [47]. The p-$FeO_2$ monolayer could potentially be synthesized through substrate-assisted growth, controlled oxidation of iron precursors, or other strategies employed in synthesizing other pentagonal 2D materials [30].

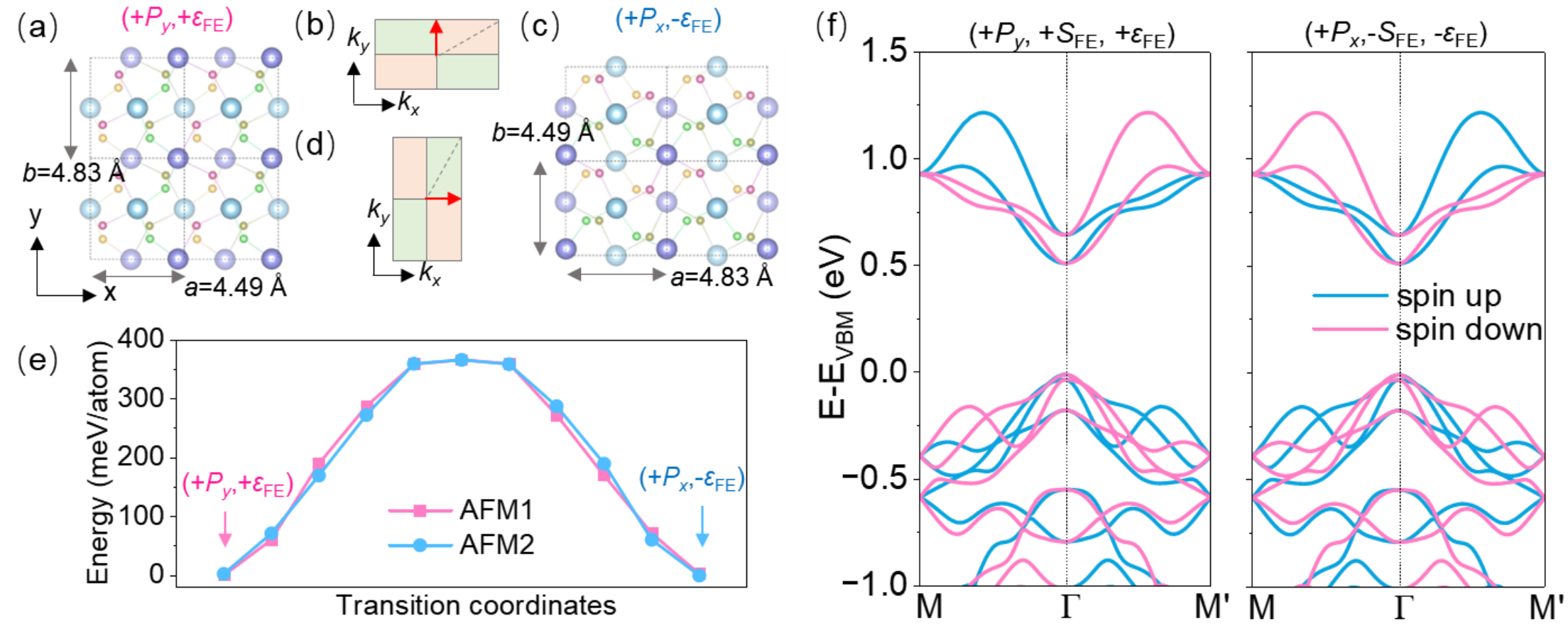

FIG. 3. (a) Top view of FE p-$FeO_2$ in the ($+P_y$, $+\varepsilon_{FE}$) state. (b) The spin splitting symmetry in the Brillouin zone in the ($+P_y$, $+\varepsilon_{AFE}$) state. The red arrows in the Brillouin zone denote the polarization directions. (c) Top view of FE p-$FeO_2$ in the ($+P_x$, $-\varepsilon_{FE}$) state. (d) The spin splitting symmetry in the Brillouin zone in the ($+P_x$, $-\varepsilon_{FE}$) state. (e) Ferroelastic transition barriers of FE p-$FeO_2$ states. (d) The spin splitting symmetry in the Brillouin zone in the ($+P_x$, $-\varepsilon_{FE}$) state. (f) Band structures of FE p-$FeO_2$ monolayer under different ferroelastic states.

The AFE-FE transition aside, the in-plane lattice anisotropy within the either FE or AFE phase indicates potential FA. The optimized lattice constants of $a$=4.49 Å and $b$=4.83 Å for the FE phase and $a$=4.35 Å and $b$=5.04 Å for the AFE phase yield reversible ferroelastic strains $\varepsilon$=(|$b$|/|$a$|-1) ×100%) of $+\varepsilon_{FE}$ = 7.6 % and $+\varepsilon_{AFE}$ =15.9 %. These values are comparable to those of other 2D ferroelastic materials, like 6.6 % in GeSe and 17.8% in GeS [57]. As a feature of FA, the sign of these values is tunable by applied in-plane strains. We replotted atomic structure shown in Fig. 1(c) in Fig. 3(a), in which electric polarization vector orients along the positive y direction ($+P_y$) and the FA strain $\varepsilon_{FE}$ is also positive ($+\varepsilon_{FE}$). The associated BZ of this structure was plotted in Fig. 3(b) where the red arrow represents the electric polarization direction in real space.

Under a uniaxial compressive strain applied along the $y$-axis, the FE p-$FeO_2$ monolayer undergoes a ferroelastic deformation that interchanges its lattice constants to $a$= 4.83Å and $b$=4.49 Å, yielding a FA strain of $-\varepsilon_{FE}$ = –7.0 %, as depicted in Fig. 3(c). The resultant lattice reshaping drives a concerted atomic displacement, rotating the spontaneous in-plane electric polarization by 90°. The polarization vector now points to the positive $x$ direction, as illustrated by the red arrow showing with the BZ schematic in Fig. 3(d). This 90°-rotation of the polarization vector exemplifies in-plane vector ferroelectricity, induced by the loss of the four-fold and two-fold rotational symmetries about the z axis (denoted operation $C_{4z}$ and $C_{2z}$). This loss of rotation axis induced in-plane polarization is analogous to the recently reported 120° vector switching that follows the loss of three-fold rotation about $z$ (operation $C_{3z}$) in monolayer $VCl_3$ [58,59]. Because both the ferroelastic and vector ferroelectric order parameters originate from the same $C_{4z}$ symmetry breaking, the p-$FeO_2$ monolayer is a directly coupled FE–FA multiferroic. The symmetry driven coupling mechanism described here can be extended to other 2D materials where $C_{3z}$ or $C_{6z}$ rotation symmetries are likewise lifted. The mentioned monolayer $VCl_3$, with its broken $C_{3z}$ symmetry, provides a concrete example.

The FA deformation surmounts an energy barrier of 366 meV/atom [Fig. 3(e)], which is among those of typical 2D ferroelastic materials, like black phosphorus (200 meV/atom) [57], AgO (400 meV/atom) [60], MnNF (336 meV/atom) [61], and $BP_5$ (336 meV/atom) [62]. A similar switching is also observed in the AFE phase (Fig. S10 [47]). In addition to the rotation of the in-plane electric polarization direction from $+y$ to $+x$, the FA switching also induces a reversal of the spin splitting. This reversal is reflected by the changed color of the 1/4 BZ marked with a dashed line, as compared between Figs. 3(b) and 3(d), and the band structures plotted in Fig. 3(f). This switching analogous to a 90° rotation of the BZ around the $z$-axis, as illustrated by Figs. 3(b) and 3(d). Therefore, the applied strain gives rise to two coupled states, namely ($+P_y$, $+S_{FE}$, $+\varepsilon_{FE}$) and ($+P_x$, $-S_{FE}$, $-\varepsilon_{FE}$), which are interrelated through ferroelastic switching, ferroelectric polarization rotation, and spin splitting reversal.

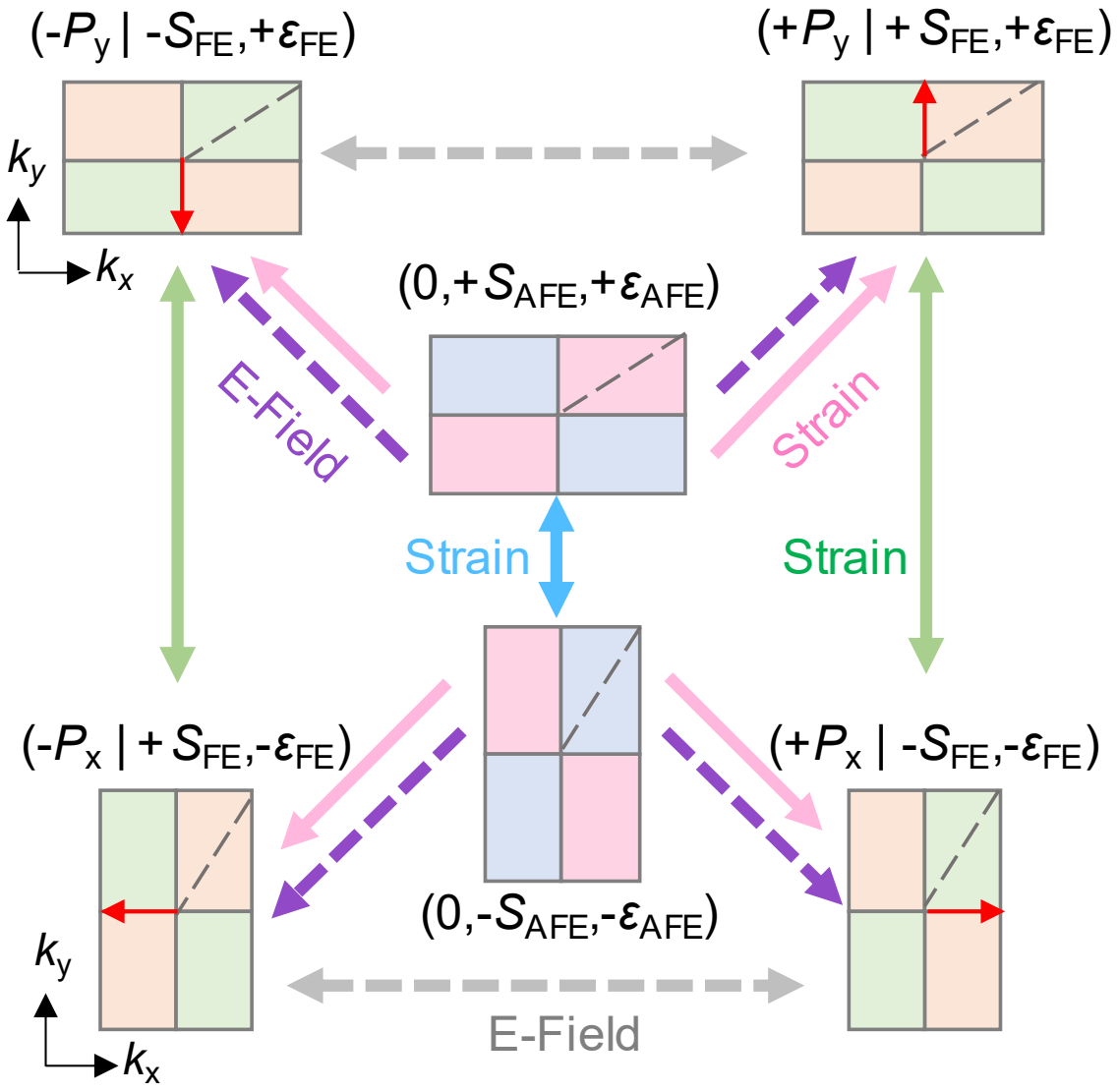

FIG. 4. Six distinct polarization states emerge from the coupling of multipe ferroic orders in the FE and AFE phases, with reversible transitions controllable via external strain or electric filed. Solid and dashed arrows denote strain-induced and electric-field-induced switching between states, respectively. Arrow colors distinguish different types of transitions. Colors inside the Brillouin zones follows the same conventions as in Fig. 2(a) and Fig. 2(b).

Considering the interplay among multiple ferroic orders in monolayer p-$FeO_2$, six distinct states (four FE and two AFE states) can be interconverted via external electric fields or strain, as illustrated in Fig. 4. As polarization $P$ are related to the combination of $S_{\mathrm{FE}}$, and $\varepsilon_{\mathrm{FE}}$, only four distinct states are independent in the FE phase, with no additional states emerging, which highlights an intrinsic coupling among $P$, $S_{FE}$, and $\varepsilon_{FE}$ order. The $S_{FE}$ and $\varepsilon_{FE}$ are uniquely determined for each polarization direction. Any change in $S_{FE}$ or $\varepsilon_{FE}$ will necessarily affect the polarization direction, while altering the polarization direction will inevitably modify the corresponding $S_{FE}$ or $\varepsilon_{FE}$ order parameters. The two AFE states exhibit opposite altermagnetic spin splitting ($S_{\mathrm{AFE}}$, blue solid arrow in Fig. 4), which can be reversed by the FA switching triggered by external in-plane strain either along the $x$ or $y$ direction. Given different lattice constants of the FE and AFE phases, in-plane strain can transform the AFE phase to the FE phase (Fig. S8 [47]), inducing substantial in-plane electric polarization in four configurations, as marked by the pink solid arrows. For instance, a compressive strain along $y$ transform the $(0, +S_{\mathrm{AFE}}, +\varepsilon_{\mathrm{AFE}})$ state to $(-P_{\mathrm{y}}|-S_{\mathrm{FE}}, +\varepsilon_{\mathrm{FE}})$ or $(+P_{\mathrm{y}}|\ +S_{\mathrm{FE}}, +\varepsilon_{\mathrm{FE}})$, in which the electric polarization direction is undetermined ($-P_{\mathrm{y}}$ or $+P_{\mathrm{y}}$) under zero electric field. In-plane electric field can more directly and determinedly switch among these states. Starting from $(0, +S_{\mathrm{AFE}}, +\varepsilon_{\mathrm{AFE}})$ state again, the structure switches to $(+P_{\mathrm{y}}|\ +S_{\mathrm{FE}}, +\varepsilon_{\mathrm{FE}})$ or $(-P_{\mathrm{y}}|-S_{\mathrm{FE}}, +\varepsilon_{\mathrm{FE}})$ under a positive or negative *y*-direction electric field (purple dashed arrows in Fig. 4), respectively. An *x*-direction electric field can transform the initial state into $(+P_{\mathrm{x}}|\ -S_{\mathrm{FE}}, -\varepsilon_{\mathrm{FE}})$ or $(-P_{\mathrm{x}}|\ +S_{\mathrm{FE}}, -\varepsilon_{\mathrm{FE}})$. For those four FE states, in-plane strain solely switches between two distinct ferroelastic states, yielding switchable AM spin splitting and reorientation of electric polarization (green solid arrows in Fig. 4). Moreover, in each group of the same ferroelastic state, with the two opposite polarization directions are interrelated through electric-field-induced switching (gray dashed arrows in Fig.4), accompanied by spin splitting reversal. In short, the AM, FE (AFE), and FA coexist and tightly couple to each other in the p-$FeO_2$ monolayer, leading to six states highly tunable by electric fields and strain.

In summary, we have theoretically demonstrated intrinsic coexistence of altermagnetic, (anti)ferroelectric, and ferroelastic orders within pentagonal monolayer $FeO_2$. Our first-principles calculations reveal two dynamically and thermodynamically stable electrically polarized phases, namely a FE and an AFE phase, exhibiting exceeding 200 K. Remarkably, we found that the spin splitting in the both phases can be effectively modulated and reversed through ferroelectric polarization switching and/or ferroelastic deformation induced by external electric fields and/or mechanical strain. Moreover, ferroelastic deformation can also directly reorient the ferroelectric polarization vector, providing a purely mechanical means to precisely control ferroelectric polarization directions. These interplays give rise to six distinct and switchable polarized states, characterized by coupled the electric polarization direction, ferroelastic strain, and altermagnetic spin splitting configuration. We also considered extensions of this structural paradigm to other transition-metal elements, like V, Cr, Mn, Co, and Ni. Although the coupled FE and FA persist, the magnetic order varies (Fig. S11 [47]), highlighting the unique altermagnetic nature of $FeO_2$. Our findings not only enrich the landscape of two-dimensional multiferroic materials but also provide promising avenues for designing future multifunctional devices exploiting couplings among mechanical, electrical, and magnetic degrees of freedom.

## Acknowledgements

We gratefully acknowledge the financial support from the National Natural Science Foundation of China (Grants No. 92477205 and No. 52461160327), the National Key R&D Program of China (Grant No. 2023YFA1406500), the Fundamental Research Funds for the Central Universities, and the Research Funds of Renmin University of China (Grants No. 22XNKJ30). Calculations were performed at the Hefei Advanced Computing Center, the Physics Lab of High-Performance Computing (PLHPC) and the Public Computing Cloud (PCC) of Renmin University of China.

## References

[1] Y. Gao, M. Gao, and Y. Lu, Two-dimensional multiferroics, Nanoscale **13**, 19324 (2021).

[2] P. Man, L. Huang, J. Zhao, and T. H. Ly, Ferroic Phases in Two-Dimensional Materials, Chem. Rev. **123**, 10990 (2023).

[3] M. M. Vopson, Fundamentals of Multiferroic Materials and Their Possible Applications, Critical Reviews in Solid State and Materials Sciences **40**, 223 (2015).

[4] N. A. Hill, Why Are There so Few Magnetic Ferroelectrics?, J. Phys. Chem. B **104**, 6694 (2000).

[5] S. Dong, H. Xiang, and E. Dagotto, Magnetoelectricity in multiferroics: a theoretical perspective, National Science Review **6**, 629 (2019).

[6] S. Dong, J.-M. Liu, S.-W. Cheong, and Z. Ren, Multiferroic materials and magnetoelectric physics: symmetry, entanglement, excitation, and topology, Advances in Physics **64**, 519 (2015).

[7] C. Wu, K. Sun, E. Fradkin, and S.-C. Zhang, Fermi liquid instabilities in the spin channel, Phys. Rev. B **75**, 115103 (2007).

[8] L. Šmejkal, J. Sinova, and T. Jungwirth, Beyond Conventional Ferromagnetism and Antiferromagnetism: A Phase with Nonrelativistic Spin and Crystal Rotation Symmetry, Phys. Rev. X **12**, 031042 (2022).

[9] S. Davoudi Tanha, M. Modarresi, and M. R. Roknabadi, Altermagnetic phase in cobalt sulfide: Effects of local correlation and crystal structure, Phys. Rev. Mater. **9**, (2025).

[10] P.-J. Guo, Y. Gu, Z.-F. Gao, and Z.-Y. Lu, Altermagnetic Ferroelectric $LiFe_2F_6$ and Spin-Triplet Excitonic Insulator Phase, arXiv:2312.13911.

[11] X. Duan, J. Zhang, Z. Zhu, Y. Liu, Z. Zhang, I. Žutić, and T. Zhou, Antiferroelectric Altermagnets: Antiferroelectricity Alters Magnets, Phys. Rev. Lett. **134**, 106801 (2025).
[12] M. Gu, Y. Liu, H. Zhu, K. Yananose, X. Chen, Y. Hu, A. Stroppa, and Q. Liu, Ferroelectric Switchable Altermagnetism, Phys. Rev. Lett. **134**, 106802 (2025).
[13] L. Šmejkal, Altermagnetic Multiferroics and Altermagnetoelectric Effect, arXiv:2411.19928.
[14] W. Sun, W. Wang, C. Yang, R. Hu, S. Yan, S. Huang, and Z. Cheng, Altermagnetism Induced by Sliding Ferroelectricity via Lattice Symmetry-Mediated Magnetoelectric Coupling, Nano Lett. **24**, 11179 (2024).
[15] W. Sun, C. Yang, W. Wang, Y. Liu, X. Wang, S. Huang, and Z. Cheng, Proposing Altermagnetic-Ferroelectric Type-III Multiferroics with Robust Magnetoelectric Coupling, Advanced Materials, 2502575 (2025).
[16] Y. Sheng, J. Liu, J. Zhang, and M. Wu, Ubiquitous van Der Waals Altermagnetism with Sliding/Moire Ferroelectricity, arXiv:2411.17493.
[17] Z. Zhu, X. Duan, J. Zhang, B. Hao, I. Žutić, and T. Zhou, Two-Dimensional Ferroelectric Altermagnets: From Model to Material Realization, Nano Lett. **25**, 9456 (2025).
[18] Z. Zhu, Y. Liu, X. Duan, J. Zhang, B. Hao, S.-H. Wei, I. Zutic, and T. Zhou, Emergent Multiferroic Altermagnets and Spin Control via Noncollinear Molecular Polarization, arXiv:2507.07039.
[19] R. Peng, S. Fang, J. Liu, and Y. S. Ang, Ferroelastic Altermagnetism, arXiv:2505.20843.
[20] S. Zhang, J. Zhou, Q. Wang, X. Chen, Y. Kawazoe, and P. Jena, Penta-graphene: A new carbon allotrope, Proceedings of the National Academy of Sciences **112**, 2372 (2015).
[21] H. L, Zhuang, From pentagonal geometries to two-dimensional materials, Computational Materials Science **159**, 448 (2019).
[22] N. A. Shah, R. A. Janjua, Y. Jin, J. Evans, J. Q. Shen, and S. He, Theoretical development in the realm of pentagonal 2D-materials, Journal of Applied Physics **136**, 130701 (2024).
[23] Y. Shen and Q. Wang, Pentagon-based 2D materials: Classification, properties and applications, Physics Reports **964**, 1 (2022).
[24] Y. Guo, C. Zhang, J. Zhou, Q. Wang, and P. Jena, Lattice Dynamic and Instability in Pentasilicene: A Light Single-Element Ferroelectric Material With High Curie Temperature, Phys. Rev. Appl. **11**, (2019).
[25] M. Debbichi and A. Alhodaib, Stability, electronic and magnetic properties of the penta-CoAsSe monolayer: a first-principles and Monte Carlo study, Phys. Chem. Chem. Phys. **24**, 5680 (2022).
[26] K. Zhao, Y. Guo, Y. Shen, Q. Wang, Y. Kawazoe, and P. Jena, Penta-BCN: A New Ternary Pentagonal Monolayer with Intrinsic Piezoelectricity, J. Phys. Chem. Lett. **11**, 3501 (2020).
[27] W. Zhang, Y. Cui, C. Zhu, B. Huang, and S. Yan, Flexible ferroelasticity in monolayer PdS2: a DFT study, Phys. Chem. Chem. Phys. **23**, 10551 (2021).
[28] W. L. Chow, P. Yu, F. Liu, J. Hong, X. Wang, Q. Zeng, C-H. Hsu, C. Zhu, J. Zhou, X. Wang, J. Xia, J. Yan, Y. Chen, D. Wu, T. Yu, Z. Shen, H. Lin, C. Jin. B. Tay, and Z. Liu, High Mobility 2D Palladium Diselenide Field-Effect Transistors with Tunable Ambipolar Characteristics, Advanced Materials **29**, 1602969 (2017).
[29] A. D. Oyedele, S. Yang, L, Liang, A. A. Puretzky, K. Wang, J. Zhang, P. Yu, P. R. Pudasaini, A. W. Ghosh, Z. Liu, C. M. Rouleau, B.G Sumpter, M. F. Chisholm, W. Zhou, P. D. Rack, D.

B. Geohegan, and K. Xiao, $PdSe_2$: Pentagonal Two-Dimensional Layers with High Air Stability for Electronics, J. Am. Chem. Soc. **139**, 14090 (2017).

[30] L. Liu, Y. Ji, M. Bianchi, S. M. Hus, Z. Li, R. Balog, J. A. Miwa, P. Hofmann, A-P Li, D. Y. Zemlyanov, Y. Li and Y. P. Chen, A metastable pentagonal 2D material synthesized by symmetry-driven epitaxy, Nat. Mater. **23**, 1339 (2024).

[31] R. Duan, C. Zhu, Q. Zeng, X. Wang, Y. Gao, Y. Deng, Y. He, J. Yang, J. Zhou, M. Xu, and Z. Liu, PdPSe: Component-Fusion-Based Topology Designer of Two-Dimensional Semiconductor, Advanced Functional Materials **31**, 2102943 (2021).

[32] J. P. Perdew, K. Burke, and M. Ernzerhof, Generalized Gradient Approximation Made Simple, Phys. Rev. Lett. **77**, 3865 (1996).

[33] P. E. Blöchl, Projector augmented-wave method, Phys. Rev. B **50**, 17953 (1994).

[34] G. Kresse and J. Furthmüller, Efficient iterative schemes for ab initio total-energy calculations using a plane-wave basis set, Phys. Rev. B **54**, 11169 (1996).

[35] G. Kresse and J. Furthmüller, Efficiency of ab-initio total energy calculations for metals and semiconductors using a plane-wave basis set, Computational Materials Science **6**, 15 (1996).

[36] S. Grimme, J. Antony, S. Ehrlich, and H. Krieg, A consistent and accurate ab initio parametrization of density functional dispersion correction (DFT-D) for the 94 elements H-Pu, The Journal of Chemical Physics **132**, 154104 (2010).

[37] Vladimir I Anisimov, F Aryasetiawan, and A I Lichtenstein, First-principles calculations of the electronic structure and spectra of strongly correlated systems: the LDA+ U method, Journal of Physics: Condensed Matter **9**, 767 (1997).

[38] A. Togo and I. Tanaka, First principles phonon calculations in materials science, Scripta Materialia **108**, 1 (2015).

[39] M. Dion, H. Rydberg, E. Schröder, D. C. Langreth, and B. I. Lundqvist, Van der Waals Density Functional for General Geometries, Phys. Rev. Lett. **92**, 246401 (2004).

[40] J. Klimeš, D. R. Bowler, and A. Michaelides, Van der Waals density functionals applied to solids, Phys. Rev. B **83**, 195131 (2011).

[41] R. D. King-Smith and D. Vanderbilt, Theory of polarization of crystalline solids, Phys. Rev. B **47**, 1651 (1993).

[42] A. Heyden, A. T. Bell, and F. J. Keil, Efficient methods for finding transition states in chemical reactions: Comparison of improved dimer method and partitioned rational function optimization method, The Journal of Chemical Physics **123**, 224101 (2005).

[43] X. Li, Z. Zhu, Q. Yang, Z. Cao, Y. Wang, S. Meng, J. Sun, and H. Gao, Monolayer puckered pentagonal VTe2: An emergent two-dimensional ferromagnetic semiconductor with multiferroic coupling, Nano Res. **15**, 1486 (2022).

[44] K. Chang, J. Liu, H. Lin, N. Wang, K. Zhao, A. Zhang, F. Jin, Y. Zhong, X. Hu, W. Duan, Q. Zhang, L. Fu, Q. Xue, X. Chen, and S-H Ji, Discovery of robust in-plane ferroelectricity in atomic-thick SnTe, Science **353**, 274 (2016).

[45] W. Wan, C. Liu, W. Xiao, and Y. Yao, Promising ferroelectricity in 2D group IV tellurides: a first-principles study, Appl. Phys. Lett. **111**, (2017).

[46] T. Zhong, X. Li, M. Wu, and J.-M. Liu, Room-temperature multiferroicity and diversified magnetoelectric couplings in 2D materials, National science review, **7**, 373-380.

[47] See Supplemental Material at XX for discussions on the anisotropic Heisenberg model,

the total energies of other magnetic configurations in the FE and AFE structures, magnetic anisotropy energy mapping, phonon spectra and AIMD simulations of p-$FeO_2$ monolayer, spin-splitting mappings, ferroelectric structural evolution, relative stability of FE and AFE phases under different lattice constants, ferroelastic transition barriers of the AFE p-$FeO_2$ monolayer and magnetic ground states of analogous structures with other transition-metal elements..

[48] H.-Y. Ma, M. Hu, N. Li, J. Liu, W. Yao, J.-F. Jia, and J. Liu, Multifunctional antiferromagnetic materials with giant piezomagnetism and noncollinear spin current, Nat Commun **12**, 2846 (2021).

[49] Y. Wu, L. Deng, X. Yin, J. Tong, F. Tian, and X. Zhang, Valley-Related Multipiezo Effect and Noncollinear Spin Current in an Altermagnet $Fe_2Se_2O$ Monolayer, Nano Lett. **24**, 10534 (2024).

[50] J. Sødequist and T. Olsen, Two-dimensional altermagnets from high throughput computational screening: Symmetry requirements, chiral magnons, and spin-orbit effects, Applied Physics Letters **124**, 182409 (2024).

[51] Y. Liu, J. Yu, and C.-C. Liu, Twisted Magnetic Van der Waals Bilayers: An Ideal Platform for Altermagnetism, Phys. Rev. Lett. **133**, 206702 (2024).

[52] L. Camerano, A. O. Fumega, J. L. Lado, A. Stroppa, and G. Profeta, Multiferroic nematic d-wave altermagnetism driven by orbital-order on the honeycomb lattice, Npj 2D Mater Appl **9**, 75 (2025).

[53] J. Wang, X. Yang, Z. Yang, J. Lu, P. Ho, W. Wang, Y. S. Ang, Z. Cheng, and S. Fang, Pentagonal 2D Altermagnets: Material Screening and Altermagnetic Tunneling Junction Device Application, Advanced Functional Materials, 2505145 (2025).

[54] Q. Liu, J. Kang, P. Wang, W. Gao, Y. Qi, J. Zhao, and X. Jiang, Inverse Magnetocaloric Effect in Altermagnetic 2D Non-van der Waals FeX (X = S and Se) Semiconductors, Adv Funct Materials **34**, 2402080 (2024).

[55] D. Guo, C. Zong, W. Zhang, C. Wang, J. Liu, and W. Ji, Tunable altermagnetism via interchain engineering in parallel-assembled atomic chains, Phys. Rev. B **112**, L041404 (2025).

[56] J. Redondo, P. Lazar, P. Procházka, S. Průša, B. Mallada, A. Cahlík, J. Lachnitt, J. Berger, B. Šmíd, L. Kormoš, P. Jelínek, J. Čechal, and M. Svec, Identification of Two-Dimensional $FeO_2$ Termination of Bulk Hematite α-$Fe_2O_3$(0001) Surface, The Journal of Physical Chemistry C **123** (23), 14312-14318. (2019).

[57] M. Wu and X. C. Zeng, Intrinsic Ferroelasticity and/or Multiferroicity in Two-Dimensional Phosphorene and Phosphorene Analogues, Nano Lett. **16**, 3236 (2016).

[58] J. Deng, D. Guo, Y. Wen, S. Lu, H. Zhang, Z. Cheng, Z. Pan, T. Jian, D. Li, H. Wang, Y. Bai, Z. Li, W. Ji, J. He, and C. Zhang, Evidence of ferroelectricity in an antiferromagnetic vanadium trichloride monolayer, Science Advances, **11**, eado6538 (2025).

[59] D. Guo, C. Wang, L. Wang, Y. Lu, H. Wu, Y. Zhang, and W. Ji, Orbital-Ordering Driven Simultaneous Tunability of Magnetism and Electric Polarization in Strained Monolayer $VCl_3$, Chinese Phys. Lett. **41**, 047501 (2024).

[60] L. Zhang, C. Tang, C. Zhang, Y. Gu, and A. Du, First-principles prediction of ferroelasticity tuned anisotropic auxeticity and carrier mobility in two-dimensional AgO, J. Mater. Chem. C **9**, 3155 (2021).

[61] M. Hu, S. Xu, C. Liu, G. Zhao, J. Yu, and W. Ren, First-principles prediction of a room-temperature ferromagnetic and ferroelastic 2D multiferroic MnNX (X = F, Cl, Br, and I), Nanoscale **12**, 24237 (2020).
[62] H. Wang, X. Li, J. Sun, Z. Liu, and J. Yang, $BP_5$ monolayer with multiferroicity and negative Poisson' s ratio: a prediction by global optimization method, 2D Mater. **4**, 045020 (2017).